\documentclass[prd,12pt,nofootinbib,superscriptaddress]{revtex4-2}
\usepackage{bm}
\usepackage{amsmath}
\usepackage{amssymb}
\usepackage{graphicx}
\usepackage{subfigure}
\usepackage{hyperref}
\usepackage{color}
\usepackage{comment}
\usepackage{ulem}
\usepackage{CJK}
\usepackage{array}
\usepackage{indentfirst}
\usepackage{float}
\hypersetup{
	colorlinks=true,
	linkcolor=red,
	citecolor=blue,
}
\begin{document}

\begin{CJK*}{UTF8}{gbsn}
\title{Importance of including higher signal harmonics in the modeling of extreme mass-ratio inspirals} 	
\author{Chao Zhang}
\email{zhangchao666@sjtu.edu.cn}
\affiliation{School of Aeronautics and Astronautics, Shanghai Jiao Tong
University, Shanghai 200240, China}
\author{Ning Dai}
\email{daining@hust.edu.cn}
\affiliation{School of Physics, Huazhong University of Science and Technology, Wuhan, Hubei
430074, China}
\author{Dicong Liang}
\email{Corresponding author. dcliang@pku.edu.cn}
\affiliation{Kavli Institute for Astronomy and Astrophysics, Peking University, Beijing
100871, China}

\begin{abstract}
Extreme mass-ratio inspirals (EMRIs) are the most potential sources detectable by the Laser Interferometer Space Antenna (LISA).
To analyze the influence of higher harmonics on parameter estimation for EMRIs efficiently, we use the waveform model that the phase trajectories are relativistic flux-based adiabatic trajectories and the waveforms are constructed by the augmented analytic kludge method.
We perform a Fisher-matrix error analysis of the EMRI parameters using signals taking into account the motion of the LISA constellation and higher harmonics of gravitational waves.
Our results demonstrate that including higher harmonics greatly reduces the errors on the exterior parameters such as inclination angle $\iota$, the luminosity distance $d_L$, the polarization angle $\psi$, and the initial phase $\Phi_0$, except for source localization $\Delta \Omega$ when EMRIs face us.
However, the influence of higher harmonics on parameters $(\iota,d_L,\psi,\Phi_0)$ can be negligible when the inclination angle is above $1.0$.
For intrinsic parameters such as the spin of central black and the masses of binaries, the influence of higher harmonics can be negligible for any inclination angle.
Our findings are independent of the mass or spin of the EMRI system.
\end{abstract}

\maketitle
\end{CJK*}

\section{Introduction}
Gravitational waves (GWs) from compact binary coalescence can be decomposed as spherical harmonic multipoles $l$
and $|m|$. In general relativity, 
the lowest polar moment of GW is the quadrupole, i.e. $(l,|m|)\, =\, (2,2)$. 
The beyond-quadrupolar multipole moments are generally referred to as higher-order modes, or higher harmonics.
For stellar-mass binary in the inspiral stage, the quadrupole mode dominates, while the higher harmonics become important when the mass ratio is extreme or the inclination angle is large.
Many waveform models have been developed to include the higher harmonics, such as Phenom family \cite{London:2017bcn, Khan:2019kot}, effective-one-body numerical relativity family \cite{Cotesta:2018fcv,  Cotesta:2020qhw, Pan:2013rra, Babak:2016tgq, Ossokine:2020kjp, Ramos-Buades:2023ehm}, and numerical relativity surrogate family \cite{Varma:2018mmi}.
Using these waveform models, significant evidence of the existence of the higher harmonics has been found in two GW events with asymmetric masses, GW190412 \cite{LIGOScientific:2020stg} and GW190814 \cite{LIGOScientific:2020zkf}.
For the event GW190521 from a heavy stellar-mass binary, 
the higher harmonics enable better constraints on the luminosity distance and inclination angle \cite{LIGOScientific:2020ufj}.
Apart from breaking the distance-inclination degeneracy, higher harmonics can also break the degeneracy between mass ratio and spin, as is shown in the study of another event from massive binary, GW170729 \cite{Chatziioannou:2019dsz}. 
The degeneracy between polarization and coalescence phase is alleviated after incorporating higher harmonics \cite{Lasky:2016knh, Payne:2019wmy}.
Furthermore, it has been investigated in many studies that systematic error is incurred from neglecting higher harmonics \cite{Varma:2014jxa, Graff:2015bba, CalderonBustillo:2015lrt, Varma:2016dnf, Littenberg:2012uj, Kalaghatgi:2019log}.

For supermassive black hole binaries (SMBHBs), the higher harmonics can be even more important.
SMBHBs are important sources for future space-based detectors, like the Laser Interferometer Space Antenna (LISA) \cite{Danzmann:1997hm, LISA:2017pwj}, TianQin \cite{TianQin:2015yph} and Taiji \cite{Hu:2017mde, Gong:2021gvw}. 
In the inspiral stage, the harmonic mode (3,3) and (4,4) can dominate the signals observed by LISA for SMBHBs with masses around $10^8~M_\odot$ \cite{Pitte:2023ltw}.
The angular resolution and distance estimation of the SMBHBs can be improved by 1 to 2 orders of magnitude after including higher harmonics \cite{Arun:2007hu, Trias:2007fp, Arun:2007qv, Porter:2008kn}.
In addition, the higher harmonics play an important role in the ringdown stage \cite{Rhook:2005pt, Berti:2005ys}. 
Although the ringdown signal is very short-lived, we can still measure the parameters to great precision by combining multiple harmonics \cite{Baibhav:2020tma, Zhang:2021kkh}.

%For most of the aforementioned studies on higher-order modes, the binary systems consist of two components with comparable masses, namely, the mass ratio is $q \gtrsim 0.1$. 
In most of the aforementioned studies on higher harmonics, the mass ratios of the binaries are comparable, which are $1 \gtrsim q \gtrsim 0.1$.
It is important and complementary to extend the study to the binary with a more extreme mass ratio $q \lesssim 10^{-4}$.
Such a binary system is called an extreme mass ratio inspiral (EMRI) system that consists of a stellar-mass compact object inspiraling into a supermassive black hole with the mass of $10^6-10^9~M_{\odot}$.
EMRI emits GWs in the millihertz band, which is one of the most potential sources for space-based detectors.
GW signals from EMRI carry highly accurate information about the sources, which enable us to test theories of gravity in strong field \cite{Yunes:2011aa, Maselli:2020zgv, Maselli:2021men, Barsanti:2022vvl, Jiang:2021htl, Zhang:2022hbt, Zhang:2022rfr, Liang:2022gdk, Zhang:2023vok} and probe the astrophysical environment around the black hole (BH) \cite{Dai:2021olt, Cardoso:2022whc, Dai:2023cft}.
Thus, it is necessary to quantify the influence of higher harmonics on the parameter estimation for EMRIs.

In this paper, we consider the EMRI system that a stellar-mass BH inspirals into a Kerr SMBH.
We obtain the evolution of the orbit by implementing the BH perturbation method and then construct the waveforms with and without higher harmonics respectively.
The errors of parameter estimation are calculated by the Fisher information matrix method.
We take LISA as a representative of the space-based detectors in our discussion,
and the analysis can be easily extended to other space-based detectors.
The paper is organized as follows. 
In Sec.~\ref{method}, we introduce the basic formalism of the BH perturbation method and calculated numerically the energy flux carried by GWs.
Then, we show the analytic post-Newtionian waveform for the quadrupole mode and higher harmonics in Sec.~\ref{waveform}.
We discuss the results of parameter estimation with the Fisher matrix in Sec.~\ref{PE}. 
Last, we summarize in Sec.~\ref{conclusion}.

\section{Method}
\label{method}
For an EMRI system composed of a small compact object with mass $m_p$ orbiting around a Kerr BH with mass $M$ and spin $a$ ($m_p\ll M$), the perturbed Einstein equations are
\begin{equation}
G^{\mu\nu}=8\pi T^{\mu\nu}_p,
\end{equation}
where
\begin{equation}
T^{\mu\nu}_p(x)=m_p\int d\tau~u^{\mu}u^\nu\frac{\delta^{(4)}\left[x-z(\tau)\right]}{\sqrt{-g}},
\end{equation}
and $u^{\mu}$ is the four-velocity of the compact object.
We study perturbations around the Kerr BH induced by the small compact object in the Newman-Penrose formalism \cite{Newman:1966ub}.
In Boyer-Lindquist coordinate,  the metric of Kerr BHs is
\begin{equation}\label{SBH}
\begin{split}
ds^2=&(1-2r/\varSigma)dt^2+(4ar\sin^2\theta/\varSigma)dtd\varphi-(\varSigma/\Delta)dr^2-\varSigma d\theta^2\\
&-\sin^2{\theta}(r^2+a^2+2a^2r\sin^2{\theta}/\varSigma)d\varphi^2.
\end{split}\end{equation}
where  $\varSigma=r^2+a^2\cos^2{\theta}$, and $\Delta=r^2-2r+a^2$. 
Based on the metric \eqref{SBH}, we construct the null tetrad,
\begin{equation}
\begin{split}
\begin{split}
l^\mu&=[(r^2+a^2)/\Delta,1,0,a/\Delta],\\
n^\mu&=[r^2+a^2,-\Delta,0,a]/(2\varSigma),\\
m^\mu&=[ia\sin{\theta},0,1,i/\sin{\theta}]/(2^{1/2}(r+ia\cos{\theta})),\\
\bar m^\mu&=[-ia\sin{\theta},0,1,-i/\sin{\theta}]/(2^{1/2}(r-ia\cos{\theta})).
\end{split}
\end{split}
\end{equation}
The propagating gravitational field is described by the complex Newman-Penrose variables
\begin{equation}
\psi_4=-C_{\alpha\beta\gamma\delta}n^\alpha \bar{m}^\beta n^\gamma \bar{m}^\delta,
\end{equation}
where $C_{\alpha\beta\gamma\delta}$ is the Weyl tensor.
A single master equation for tensor ($s=-2$) perturbations was derived as \cite{Teukolsky:1973ha},

\begin{equation}
\label{TB}
\begin{split}
&\left[\frac{(r^2+a^2)^2}{\Delta}-a^2\sin^2{\theta}\right]\frac{\partial^2\psi}{\partial t^2}+\frac{4ar}{\Delta}\frac{\partial^2\psi}{\partial t\partial\varphi}+\left[\frac{a^2}{\Delta}-\frac{1}{\sin^2{\theta}}\right]\frac{\partial^2\psi}{\partial \varphi^2}\\
&\qquad-\Delta^{-s}\frac{\partial}{\partial r}\left(\Delta^{s+1}\frac{\partial\psi}{\partial r}\right)-\frac{1}{\sin\theta}\frac{\partial}{\partial \theta}\left(\sin\theta\frac{\partial\psi}{\partial \theta}\right)-2s\left[\frac{a(r-1)}{\Delta}+\frac{i\cos\theta}{\sin^2{\theta}}\right]\frac{\partial\psi}{\partial \varphi}\\
&\qquad\qquad\qquad\qquad-2s\left[\frac{(r^2-a^2)}{\Delta}-r-ia\cos\theta\right]\frac{\partial\psi}{\partial t}+(s^2\cot^2\theta-s)\psi=4\pi\varSigma T,
\end{split}
\end{equation}
the explicit field $\psi=(r-i a\cos\theta)^{4}\psi_4$ and the corresponding source $T$ are given in \cite{Teukolsky:1973ha}.
In terms of the eigenfunctions ${_{s}}S_{lm}(\theta)$ \cite{Teukolsky:1973ha,Goldberg:1966uu}, the field $\psi$ can be written as   %
\begin{equation}
\psi=\int d\omega \sum_{l,m}R_{\omega lm}(r)~{_{s}}S_{lm}(\theta)e^{-i\omega t+im\varphi},
\end{equation}
where the radial function $R_{\omega lm}(r)$ satisfies the inhomogeneous Teukolsky equation
\begin{equation}
\label{Teukolsky}
\Delta^{-s}\frac{d}{d r}\left(\Delta^{s+1}\frac{d R_{\omega lm}}{d r}\right)-V_{T}(r)R_{\omega lm}=T_{\omega lm},
\end{equation}
%the function
and
\begin{equation}
V_{T}=-\frac{K^2-2is(r-1)K}{\Delta}-4is\omega r+\lambda_{lm\omega},
\end{equation}
 $K=(r^2+a^2)\omega-am$, $\lambda_{lm\omega}$ is the corresponding eigenvalue which can be computed by using the BH Perturbation Toolkit \cite{BHPToolkit}, and the source $T_{\omega lm}(r)$ is
\begin{equation}
T_{\omega lm}(r)=\frac{1}{2\pi}\int dt d\Omega ~4\pi \Sigma T ~{_s}S_{lm}(\theta)e^{i\omega t-im\varphi}.
\end{equation}
The homogeneous Teukolsky equation \eqref{Teukolsky} admits two linearly independent solutions $R^{\text{in}}_{\omega lm}$ and $R^{\text{up}}_{\omega lm}$, with the following asymptotic values at the horizon $r_+$ and at infinity,
\begin{equation}
R^{\text{in}}_{\omega lm}=
\begin{cases}
B^{\text{tran}}\Delta^{-s}e^{-i\kappa r^*},&\quad (r\to r_+)\\
B^{\text{out}}\frac{e^{i\omega r^*}}{r^{2s+1}}+B^{\text{in}}\frac{e^{-i\omega r^*}}{r}, &\quad (r\to+\infty)
\end{cases}
\end{equation}
\begin{equation}
R^{\text{up}}_{\omega lm}=
\begin{cases}
D^{\text{out}}e^{i\kappa r^*}+\frac{D^{\text{in}}}{\Delta^{s}}e^{-i\kappa r^*},&\quad (r\to r_+)\\
D^{\text{tran}}\frac{e^{i\omega r^*}}{r^{2s+1}},&\quad (r\to+\infty)\\
\end{cases}
\end{equation}
where $\kappa=\omega-m a/(2r_+)$, $r_\pm=1\pm\sqrt{1-a^2}$, and the tortoise radius of the Kerr metric
\begin{equation}
r^*=r+\frac{2r_+}{r_+-r_-}\ln \frac{r-r_+}{2}-\frac{2r_-}{r_+-r_-}\ln \frac{r-r_-}{2}.
\end{equation}
With the help of these homogeneous solutions, the solution to Eq.~\eqref{Teukolsky} is

\begin{equation}
\begin{split}
\label{Rwlm(r)}
R_{\omega lm}(r)=\frac{1}{W}
\left(R^{\text{in}}_{\omega lm}\int_{r}^{+\infty}\Delta^{s}R^{\text{up}}_{\omega lm}T_{\omega lm}dr+R^{\text{up}}_{\omega lm}\int_{r_+}^{r}\Delta^{s}R^{\text{in}}_{\omega lm}T_{\omega lm}dr\right),
\end{split}
\end{equation}
and the constant Wronskian given by
\begin{equation}
W= \Delta^{s+1} \left(R^{\text{in}}_{\omega lm}\frac{d R^{\text{up}}_{\omega lm}}{dr}-R^{\text{up}}_{\omega lm}\frac{d R^{\text{in}}_{\omega lm}}{dr}\right)=2i\omega B^{\text{in}} D^{\text{tran}}.
\end{equation}
The solution in Eq. \eqref{Rwlm(r)} is purely outgoing at infinity and purely ingoing at the horizon,
\begin{equation}
\begin{split}
R_{\omega lm}(r\to r_+)=Z^{\infty}_{\omega lm}\Delta^{-s}e^{-i\kappa r^*},\\
R_{\omega lm}(r\to \infty)=Z^{H}_{\omega lm}r^{-2s-1}e^{i\omega r^*},
\end{split}
\end{equation}
with
\begin{equation}
\begin{split}
Z^{\infty}_{\omega lm}&=\frac{B^{\text{tran}}}{W}\int_{r_+}^{+\infty}\Delta^{s}R^{\text{up}}_{\omega lm}T_{\omega lm}dr,\\
Z^{H}_{\omega lm}&=\frac{D^{\text{tran}}}{W}\int_{r_+}^{+\infty}\Delta^{s}R^{\text{in}}_{\omega lm}T_{\omega lm}dr.
\label{amplitudes}
\end{split}
\end{equation}
For a circular equatorial orbit with orbital angular frequency $\hat{\omega} $, we get
\begin{equation}
Z^{H,\infty}_{\omega lm}=\delta(\omega-m \hat{\omega})\mathcal{A}^{H,\infty}_{\omega lm}.
\end{equation}
The gravitational energy fluxes at infinity and at the horizon are respectively given by
\begin{equation}
\begin{split}
\dot{E}_{\text{grav}}^{\infty}=\left(\frac{d E}{dt}\right)_{\text{grav}}^\infty&=\sum_{l=2}^{\infty}\sum_{m=1}^{l}\frac{|\mathcal{A}^{H}_{\omega lm}|^2}{2\pi\omega^2}, \\
\dot{E}_{\text{grav}}^H=\left(\frac{d E}{dt}\right)_{\text{grav}}^H&=\sum_{l=2}^{\infty}\sum_{m=1}^{l}\alpha^G_{lm}\frac{|\mathcal{A}^{\infty}_{\omega lm}|^2}{2\pi\omega^2},
\end{split}
\end{equation}
where the coefficient $\alpha^G_{l m}$ is \cite{Hughes:1999bq}
\begin{equation}\label{energyformula2}
\alpha^G_{l m}=\frac{256\left(2 r_{+}\right)^5 \kappa\left(\kappa^2+4 \epsilon^2\right)\left(\kappa^2+16 \epsilon^2\right)\omega^3}{\left|B_G\right|^2},
\end{equation}
and
\begin{equation}
\begin{aligned}
\left|B_G\right|^2 &=\left[\left(\lambda_{l m \omega}+2\right)^2+4 a\omega-4 a^2\omega^2\right]\times\left[\lambda_{l m \omega}^2+36 m a\omega-36 a^2\omega^2\right] \\
&+\left(2 \lambda_{l m \omega}+3\right)\left[96 a^2\omega^2-48 m a\omega\right]+144\omega^2\left(1-a^2\right) .
\end{aligned}
\end{equation}
Therefore, the total energy fluxes emitted from the EMRIs read
\begin{equation}
\label{e-total}
\dot{E}_{\text{grav}}=\dot{E}_{\text{grav}}^\infty+\dot{E}_{\text{grav}}^H.
\end{equation}
The energy flux emitted by tensor fields can also be computed with the BH Perturbation Toolkit \cite{BHPToolkit}.
Considering the circular equatorial trajectory at $r_0$,
the sources are
\begin{equation}
T^{\mu\nu}_p(x)=\frac{m_p}{r_0^2}\frac{u^{\mu}u^{\nu}}{u^t}\delta(r-r_0)\delta(\cos\theta)\delta(\varphi-\hat{\omega} t),
\end{equation}
where $\hat{\omega}$ is the orbital angular frequency.
There are three constants for the geodesic motion in Kerr spacetime,
which are the specific energy $\hat{E}$, the angular momentum $\hat{L}$, and the Carter constant $\hat{Q}$.
The geodesic equations are 
\begin{eqnarray}
m_p\Sigma \frac{d t}{d \tau} &=&\hat{E} \frac{\varpi^{4}}{\Delta}+a \hat{L}\left(1-\frac{\varpi^{2}}{\Delta}\right)-a^{2} \hat{E}\sin^{2}\theta, \label{timequa}\\
m_p\Sigma \frac{d r}{d \tau} &=&\pm \sqrt{V_{r}\left(r_{0}\right)}, \label{vrqua}\\
m_p\Sigma \frac{d \theta}{d \tau} &=&\pm \sqrt{V_{\theta}\left(\theta\right)}, \label{vthetaqua}\\
m_p\Sigma \frac{d \varphi}{d \tau} &=&a \hat{E}\left(\frac{\varpi^{2}}{\Delta}-1\right)-\frac{a^{2} \hat{L}}{\Delta}+ \hat{L} \csc ^{2} \theta,\label{anglequa}
\end{eqnarray}
where $\varpi\equiv\sqrt{r^2+a^2}$, the radial and polar potentials are
\begin{eqnarray}
&V_{r}(r)& =\left(\hat{E} \varpi^{2}-a \hat{L}\right)^{2}-\Delta\left(r^{2}+\left(\hat{L}-a \hat{E}\right)^{2}+\hat{Q}\right), \\
&V_{\theta}(\theta)& = \hat{Q}-\hat{L}^{2} \cot ^{2} \theta-a^{2}\left(1-\hat{E}^{2}\right) \cos ^{2} \theta.
\end{eqnarray}
For a quasi-circular orbit on the equatorial plane,
% the coordinates $r$ and $\theta$ are considered constants in the adiabatic approximation.
% Then Eqs. \eqref{vrqua} and \eqref{vthetaqua} are vanished, Eqs. \eqref{timequa}  and \eqref{anglequa} are remained.
the conserved constants are \cite{Detweiler:1978ge}
\begin{eqnarray}
    \label{orbitE}
	\hat{E}&=& m_p\frac{r_0^{3 / 2}-2 r_{0}^{1 / 2} \pm a }{r_{0}^{3 / 4}\left(r_{0}^{3 / 2}-3  r_{0}^{1 / 2} \pm 2 a \right)^{1 / 2}},    \\
	\hat{L}&=&m_p \frac{\pm (r_{0}^{2}\mp 2a r_{0}^{1 / 2} + a^2)}{r_{0}^{3 / 4}\left(r_{0}^{3 / 2}-3  r_{0}^{1 / 2} \pm 2 a \right)^{1 / 2}},\\
	\hat{Q}&=&0.
\end{eqnarray}
The orbital angular frequency is
\begin{equation}\label{orbitF}
\hat{\omega} \equiv \frac{d\varphi}{dt}=\frac{\pm 1}{r_{0}^{3 / 2} \pm a},
\end{equation}
where $\pm$ corresponds to  co-rotating (+) or counter-rotating (-).
In the following discussions, we only consider the co-rotating cases.

\section{Waveform including higher harmonics}
\label{waveform}
For EMRIs, the gravitational radiation reaction acting on the massive particle can be split into two parts: the dissipative one and the conservative one.
When the evolution time of the geodesic motion is much longer than the time scale of the orbital period,
we can consider the dissipative part under the adiabatic approximation only.
The dissipative part can be calculated from the energy flux at the horizon of the central BH and at infinity.
Assuming the motion is strictly geodesic over several orbital periods,
we calculate the energy flux from the system and then give the time-averaged rates of change of the orbital parameters.
Combining Eq. \eqref{e-total} and Eq. \eqref{orbitE}, the energy balance equation is 
\begin{equation}
\label{balance}
   \dot{E}_{\text{grav}}= \left\langle \frac{d E_{\text{grav}}}{dt} \right \rangle=-m_p\frac{d\hat{E}}{dt}.
\end{equation}
As pointed out in Ref. \cite{Cutler:1994pb},
as long as the extreme mass ratio is suitably satisfied,
the results are compatible with the initial assumption, and the calculation is self-consistent.
The orbital evolution is determined by
 \begin{equation}\label{orbittime}
 \frac{d r}{dt}=-\frac{\dot{E}_{\text{grav}}}{m_p}\left(\frac{d \hat{E}}{dr}\right)^{-1},\qquad \frac{d \varphi_{\text{orb}}}{d t}=\pi f,
 \end{equation}
 $f=\hat{\omega}/\pi$ is the GW frequency.
 We can obtain the inspiral trajectory from adiabatic evolution in Eq. \eqref{orbittime}, then compute GWs in the post-Newtonian expansion \cite{Poisson:1993vp, Fujita:2010xj, Shibata:1994jx, Sasaki:2003xr}.
 The waveforms in the quadrupole formula can be obtained from $(l=2, m=2)$,
\begin{equation}\label{mode22}
\begin{split}
h_+^{(2,2)}&=\mathcal{A}\left(\cos^2\iota +1\right) \cos [2\varphi_{\text{orb}}(t)+2\Phi_0],\\
h_\times^{(2,2)} &= -2\mathcal{A} \cos \iota  \sin [2\varphi_{\text{orb}}(t)+2\Phi_0],
\end{split}
\end{equation}
while higher harmonics including $(l=2,m=1)$ and $(l=3,m=3)$ are
\begin{equation}\label{mode33}
\begin{split}
h_+^{(2,1)}+h_+^{(3,3)}&=\mathcal{A}[M\hat{\omega}(t)]^{1/3}\frac{\sin \iota }{8}  \left(\left(\cos ^2\iota +5\right) \sin [\varphi_{\text{orb}}(t)+\Phi_0]\right.\\
&\left.\qquad\qquad+9 \left(\cos ^2\iota +1\right) \sin [3\varphi_{\text{orb}}(t)+3\Phi_0]\right),\\
h_\times^{(2,1)}+h_\times^{(3,3)}&=\mathcal{A}[M\hat{\omega}(t)]^{1/3}\frac{3\sin (2\iota ) }{8} \left(\cos [\varphi_{\text{orb}}(t)+\Phi_0]+3 \cos [3\varphi_{\text{orb}}(t)+\Phi_0]\right),
\end{split}
\end{equation}
and higher harmonics including $(l=3,m=2)$ and $(l=4,m=4)$ are
\begin{equation}
    \begin{split}
     h_+^{(3,2)}+h_+^{(4,4)}&=-\mathcal{A}[M\hat{\omega}(t)]^{2/3}\frac{1}{6}\left(\left(19+9\cos^2\iota-2\cos^4\iota\right)\cos[2\varphi_{\text{orb}}(t)+2\Phi_0]\right.\\
     &\left.\qquad\qquad +8\sin^2\iota\left(1+\cos^2\iota\right)\cos[4\varphi_{\text{orb}}(t)+4\Phi_0]\right),\\
 h_\times^{(3,2)}+h_\times^{(4,4)}&=\mathcal{A}[M\hat{\omega}(t)]^{2/3}\frac{1}{3}\left(\left(17-4\cos^2\iota\right)\sin[2\varphi_{\text{orb}}(t)+2\Phi_0] \right.\\
 &\left.\qquad\qquad+8\sin^2\iota\sin[4\varphi_{\text{orb}}(t)+4\Phi_0]\right),
    \end{split}
\end{equation}
where $\iota$ is the inclination angle between the binary orbital angular momentum and the line of sight, $\Phi_0$ is the initial phase, $\mathcal{A}=2m_{\rm p}\left[M\hat{\omega}(t)\right]^{2/3}/d_L$ is the GW amplitude and $d_L$ is the luminosity distance of the source.
Introducing the polarization angle $\psi$, the polarizations $h_+$ and $h_\times$ transform according to 
\begin{equation}
h_+\to \cos (2\psi) h_++\sin (2\psi) h_\times,\qquad h_\times\to -\sin (2\psi) h_++\cos (2\psi) h_\times. 
\end{equation}
The GW strain measured by the detector is
\begin{equation}\label{signal}
h(t)=h_{+}(t) F^{+}(t)+h_{\times}(t) F^{\times}(t),
\end{equation}
where the interferometer pattern functions $F^{+,\times}(t)$ can be expressed in terms of the source orientation $(\theta_s,\phi_s)$.
Finally, the GW signals are modulated due to the LISA orbital motion \cite{Babak:2006uv}.
We account for this effect by modifying the phase as
\begin{equation}\label{doppler}
\varphi_{\text{orb}}(t)\to \varphi_{\text{orb}}(t)+\varphi_{\text{orb}}'(t)R_{\text{AU}}\sin\theta_s\cos\left(\frac{2\pi t}{T} -\phi_s-\phi_{\alpha}\right),
\end{equation}
where $\phi_{\alpha}$ is the ecliptic longitude of the detector $\alpha$ at $t=0$,
the rotational period $T$ is 1 year and the radius of the orbit $R_{\text{AU}}$ is 1 AU.
The signals \eqref{signal} measured by the detector are determined by the following ten parameters
\begin{equation}
\xi=(\ln M, \ln m_p, a, r_0, \Phi_0, \theta_s, \phi_s, \iota, \psi, d_L),
\end{equation}
where $r_0$ is the initial orbital separation.
The signal-to-noise ratio (SNR) of the GW signals is 
\begin{equation}
\rho=\sqrt{\left\langle h|h \right\rangle},
\end{equation}
the noise-weighted inner product between two templates $h_1$ and $h_2$ is
\begin{equation}\label{product}
\left\langle h_{1} \mid h_{2}\right\rangle=4 \Re \int_{f_{\min }}^{f_{\max }} \frac{\tilde{h}_{1}(f) \tilde{h}_{2}^{*}(f)}{S_{n}(f)} df,
\end{equation}
where
\begin{equation}
    f_{\text{max}}=\text{min}(f_{\text{ISCO}},f_{\text{up}}),~~~~~~f_{\text{min}}=\text{max}(f_{\text{low}},f_{\text{start}}),
\end{equation}
$f_{\text{ISCO}}$ is the GW frequency at the inner-most stable circular orbit (ISCO) \cite{Jefremov:2015gza},
$f_{\text{start}}$ is the initial frequency at $t=0$,
the cutoff frequencies $f_{\text{low}}=10^{-4}$ Hz and $f_{\text{up}}=1$ Hz, $\tilde{h}(f)$ is the Fourier transform of the time-domain signal $h(t)$,
its complex conjugate is $\tilde{h}^{*}(f)$,
and $S_n(f)$ is the noise spectral density of the space-based GW detectors.

 In the large SNR limit,
the posterior probability distribution of the source parameters $\xi$ can be approximated by a multivariate Gaussian distribution centered around the true values $\hat{\xi}$.
Assuming flat or Gaussian priors on the source parameters $\xi$,
their covariances are given by the inverse of the Fisher information matrix (FIM)
\begin{equation}
\Gamma_{i j}=\left\langle\left.\frac{\partial h}{\partial \xi_{i}}\right| \frac{\partial h}{\partial \xi_{j}}\right\rangle_{\xi=\hat{\xi}}.
\end{equation}
The statistical error $\sigma_{\xi_i}$ on $\xi_i$ and the correlation coefficients $c_{\xi_{i} \xi_{j}}$ between the parameters are provided by the diagonal and non-diagonal parts of ${\bf \Sigma}={\bf \Gamma}^{-1}$, i.e.
\begin{equation}
\sigma_{\xi_i}=\Sigma_{i i}^{1 / 2} \quad, \quad c_{\xi_{i} \xi_{j}}=\Sigma_{i j} /\left(\sigma_{\xi_{i}} \sigma_{\xi_{j}}\right),
\end{equation}
and the angular uncertainty of the sky localization is evaluated as \cite{Cutler:1997ta}
\begin{equation}
\Delta \Omega\equiv2\pi\sin\theta_s
\sqrt{\sigma_{\theta_s\theta_s}\sigma_{\phi_s\phi_s}-\sigma^2_{\theta_s\phi_s}}\,.
\end{equation}
Because of the triangle configuration of the space-based GW detector that can be considered as two L-shape detectors, 
so the total SNR is defined by $\rho=\sqrt{\rho_1^2+\rho_2^2}$ and the total covariance matrix of the binary parameters is obtained by inverting the sum of the Fisher matrices $\bf \Sigma=(\Gamma_1+\Gamma_2)^{-1}$ \cite{Cutler:1997ta}.

\section{Parameter Estimation Errors}
\label{PE}

In this section, we analyze the influence of higher harmonics on the errors of source parameters discussed in the previous section.
For small inclination $\iota$, the GW polarizations $h_+$ and $h_\times$ in Eq.~\eqref{mode22} can be expanded 
\begin{equation}
\begin{split}
h_+^{(2,2)}&=2m_{\rm p}\left[M\hat{\omega}(t)\right]^{2/3}\frac{2-\iota ^2 }{d_L}\cos [2\varphi_{\text{orb}}(t)+2\Phi_0+2\psi],\\
h_\times^{(2,2)} &= -2m_{\rm p}\left[M\hat{\omega}(t)\right]^{2/3}\frac{2-\iota ^2 }{d_L} \sin [2\varphi_{\text{orb}}(t)+2\Phi_0+2\psi].
\label{HModesEQ}
\end{split}
\end{equation}
We notice that $\iota$ and $d_L$ are highly correlated with each other in a small inclination angle.
Similarly, the strong degeneracy of parameters $\psi$ and $\Phi_0$ for the small inclination angle exists.
The above degeneracies  will deteriorate the ability to measure source parameters for space-based detectors.
The higher harmonics in Eq.~\eqref{mode33} in small inclination approximation are
\begin{equation}
\begin{split}
h_+^{(2,1)}+h_+^{(3,3)}&=m_{\rm p}\left[M\hat{\omega}(t)\right]\frac{3\iota}{2d_L} \left(\sin [\varphi_{\text{orb}}(t)+\Phi_0-2\psi]+3\sin [3\varphi_{\text{orb}}(t)+3\Phi_0-2\psi]\right),\\
h_\times^{(2,1)}+h_\times^{(3,3)}&=m_{\rm p}\left[M\hat{\omega}(t)\right]\frac{3\iota}{2d_L}\left(\cos [\varphi_{\text{orb}}(t)+\Phi_0-2\psi]+3 \cos [3\varphi_{\text{orb}}(t)+3\Phi_0-2\psi]\right).
\end{split}
\end{equation}
As we can see, higher harmonics have different dependence on inclination and initial phase compared with Eqs.~\eqref{HModesEQ}, so including higher harmonics can break some of the degeneracies that currently haunt the parameter estimation in small inclinations.
For example, the distance-inclination degeneracy, the degeneracy between the initial phase and polarization angle can be broken by adding the higher harmonics GW signal.
Thus including higher harmonics may greatly reduce the errors of parameters $(\iota,d_L,\psi,\Phi_0)$ in small inclinations, but the impact on intrinsic parameters $(M, m_p, a)$ and the source localization still needs to be studied.
Another question is whether higher harmonics still play an important role in parameter estimation when the small-inclination condition breaks?
In order to estimate the influence of higher harmonics on the errors of the source parameters, the FIM method is applied for LISA to numerically calculate the estimation errors on kinds of simulated EMRI signals including higher harmonics.

\begin{figure*}[htp]
  \centering
  \includegraphics[width=0.9\textwidth]{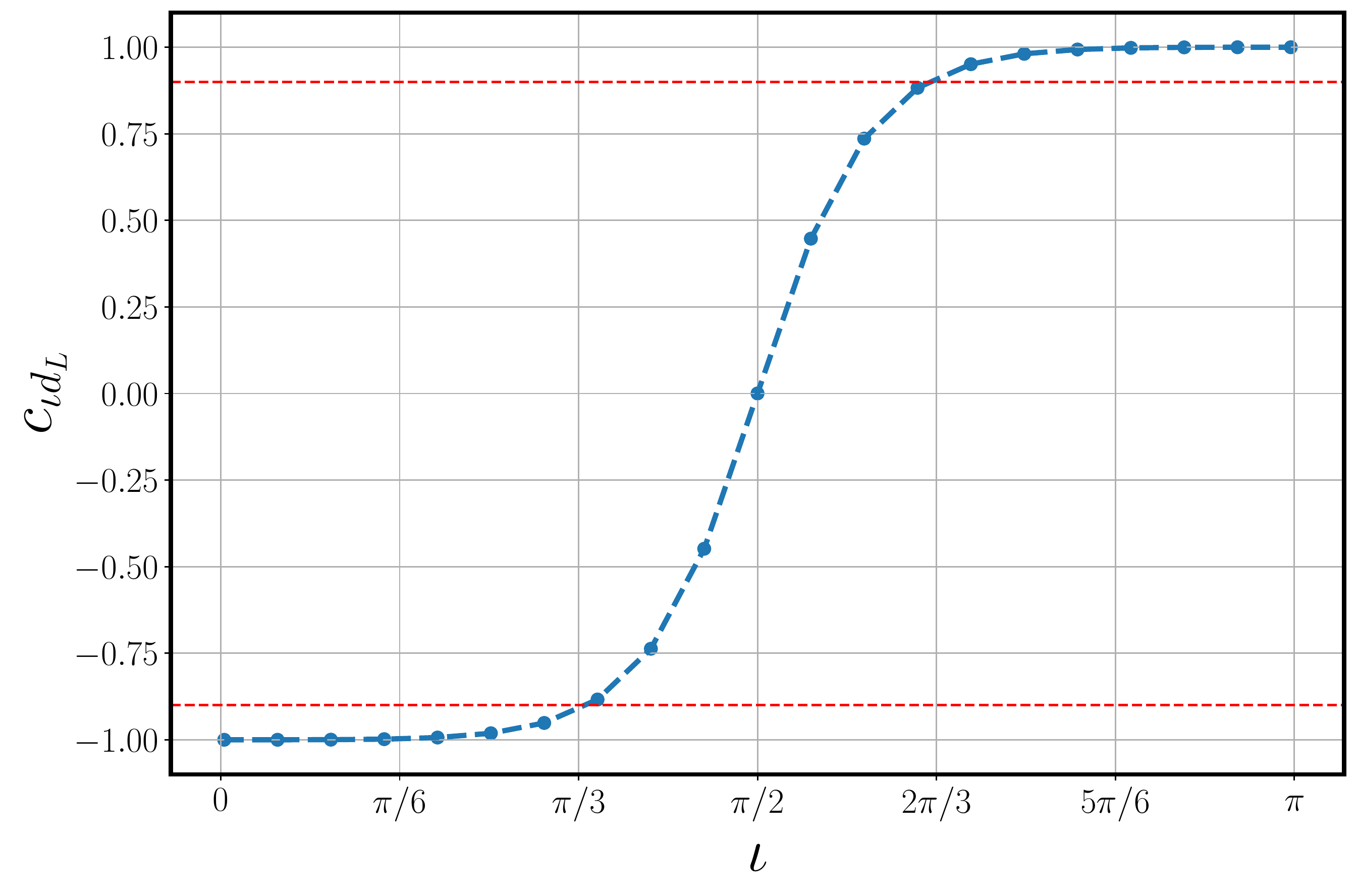}
  \caption{The correlation coefficient $c_{\iota d_L}$ between the inclination angle $\iota$ and the luminosity distance $d_L$ as functions of inclination angle.
  The blue dash line represents the medians of the correlation coefficient $c_{\iota d_L}$ of the samples in which the GW waveform only contains mode $(l,m)=(2,2)$. The horizontal red dashed line represents the correlation coefficient whose magnitude is $0.9$.
  }
  \label{coeff}
\end{figure*}

We choose EMRI systems with mass $M=10^6~M_\odot$ and dimensionless spin $a=0.9$ for the central BH, and $m_p=10~M_\odot$ for the small BH.
The initial phase is set as $\Phi_0=0$ and the initial orbital separation $r_0$ is adjusted to experience one-year adiabatic evolution before the plunge $r_{\text{end}}=r_{\text{ISCO}}+0.1~M$.
The luminosity distance $d_L$ is set to be $1$ Gpc.
To prove our results general and independent of the particular choice of the source location, we simulate 1000 sources with parameters ($\cos\theta_s,\phi_s,\psi_s$) uniformly distributed for each inclination $\iota$ value. 
The inclination angle is chosen uniformly in $[0,2\pi]$.
Figure \ref{coeff} shows the medians of the correlation coefficient $c_{\iota d_L}$ between the inclination angle $\iota$ and the luminosity distance $d_L$ when we use GW waveform only from the mode $(l,m)=(2,2)$.  
In general, for two parameters that are very highly correlated with each other, the magnitude of the correlation coefficient is large $|c|\gtrsim 0.9$ \cite{schober_correlation_2018}.
As the inclination angle $\iota$ increases from $0$ to $\pi$, the correlation coefficient $c_{\iota d_L}$ increases from $-1$ to $1$.
The inclination angle $\iota$ and the luminosity distance $d_L$ are very highly anti-correlated with each other when the inclination angle $\iota \lesssim 1$ while they are very highly correlated with each other when the inclination angle $\iota \gtrsim \pi-1$. 
The opposite correlation between $\iota$ and $d_L$ at $\iota \lesssim 1$ and $\iota \gtrsim \pi-1$ is due to the fact that the GW waveform is invariant under the transformation from $\iota$ to $\pi-\iota$ in Eq.~\eqref{mode22}.
Furthermore, the magnitude of correlation coefficient $c_{\iota d_L}$ increases as the inclination angle decreases from $\pi/2$ to $0$.
Especially, the magnitude of correlation coefficient $c_{\iota d_L}$ reaches 1 when the inclination angle is near $0$.
As we have discussed before, the GW waveform for the plus mode $h_+$ and the cross mode $h_{\times}$ in small inclination expansion can be expressed as Eq.~\eqref{HModesEQ}.
We can see that $h_+\propto (2-\iota ^2)/d_L$ and $h_\times\propto (2-\iota ^2)/d_L$, then the total strain is also $h(t)\propto (2-\iota ^2)/d_L$, thus, the inclination angle and luminosity distance are highly correlated with each other for small inclination.

\begin{figure*}[htp]
  \centering
  \includegraphics[width=0.9\textwidth]{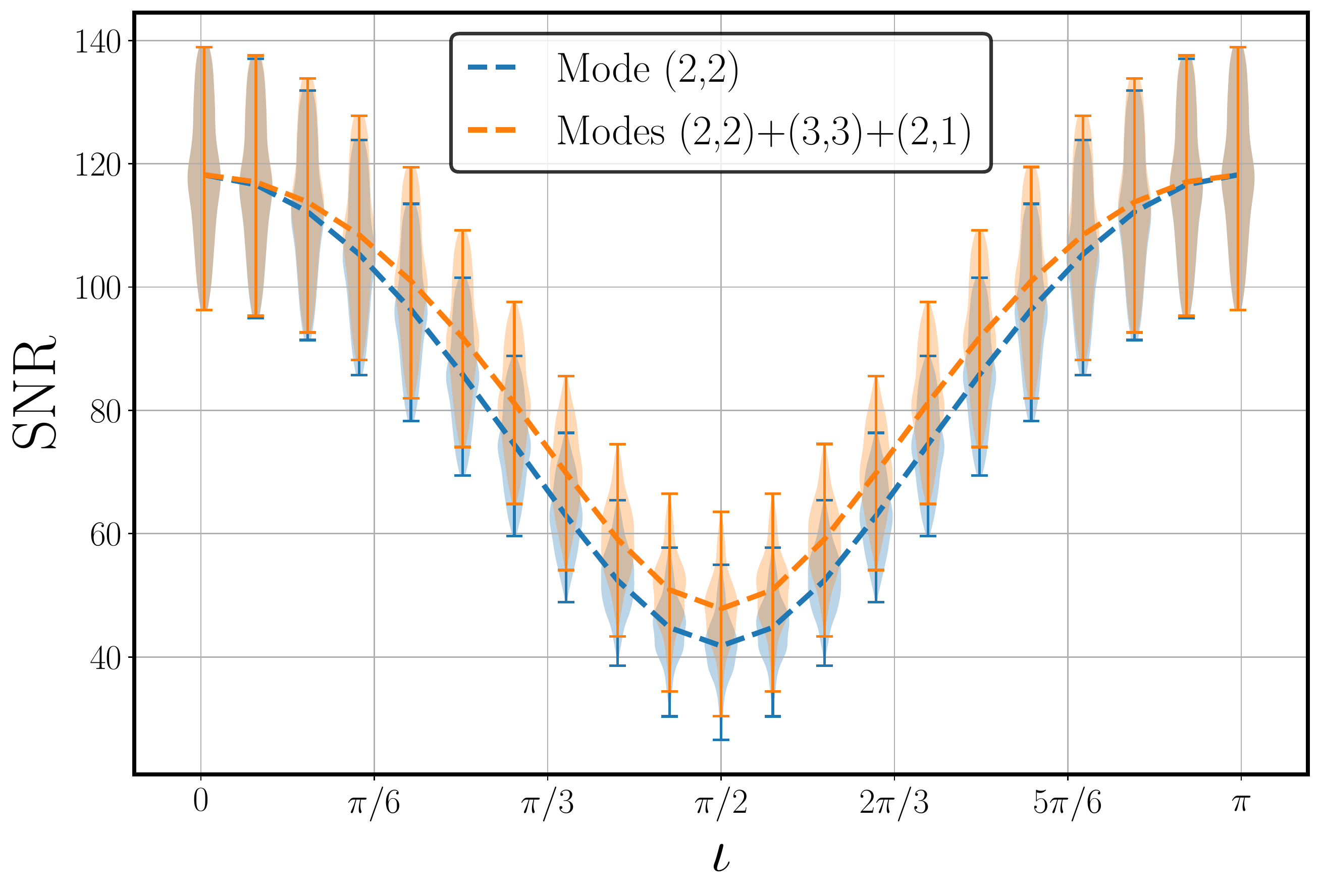}
  \caption{The SNR as functions of inclination angle.
  The blue dash line represents the medians of SNR of the samples in which the GW waveform only contains mode $(l,m)=(2,2)$, while the orange dash line for those including higher harmonics modes $(l,m)=(2,2),(2,1),(3,3)$. The distribution of SNR of the samples is denoted with the violin plots.
  }
  \label{snr1}
\end{figure*}

\begin{figure*}[htp]
  \centering
  \includegraphics[width=0.9\textwidth]{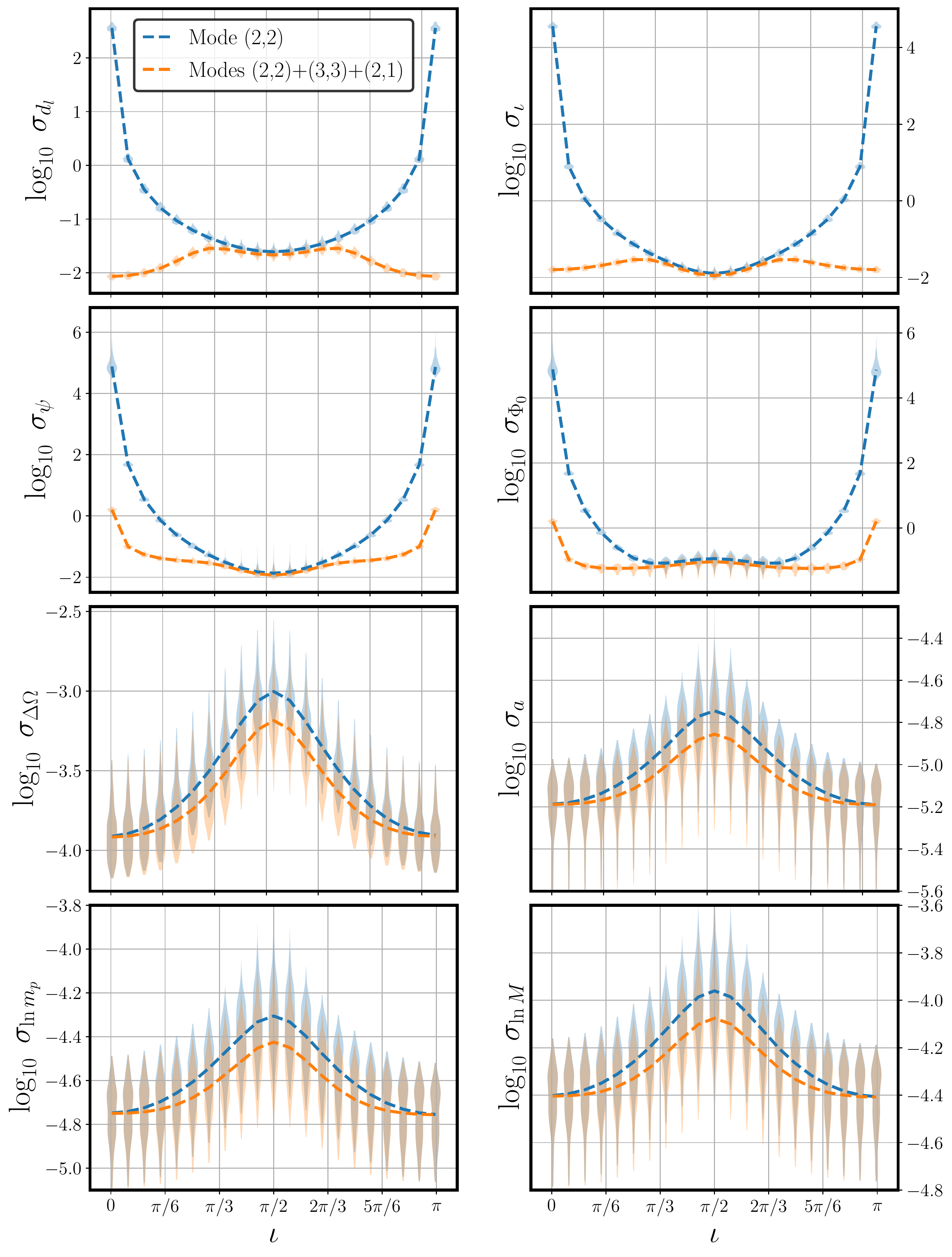}
  \caption{The medians of parameter estimation errors as functions of inclination angle.
  The blue dash line represents the medians of errors of the samples in which the GW waveform only contains mode $(l,m)=(2,2)$, while the orange dash line for those including higher harmonics modes $(l,m)=(2,2),(2,1),(3,3)$. The distribution of errors in the samples is denoted with the violin plots.
  }
  \label{FIMresult1}
\end{figure*}

Figures \ref{snr1} and \ref{FIMresult1} respectively give the medians and the distribution of SNR and the parameter estimation errors  for each inclination value when we use GW waveform only from  mode $(l,m)=(2,2)$  without higher harmonics, as well as the GWs including higher harmonics modes $(l,m)=(2,2),(2,1),(3,3)$.

For GW waveform only from mode $(l,m)=(2,2)$ without higher harmonics, we notice strong dependence of $\sigma_{\iota}$, $\sigma_{d_L}$, $\sigma_{\psi}$ and $\sigma_{\Phi_0}$ on the inclination angle and the higher uncertainty in the measurement of the astrophysical parameters for the smaller inclination angle.
For intrinsic parameters $(M, m_p, a)$ and the source localization $\Delta \Omega$, the behaviors of errors are contrast with parameters $(\iota,d_L,\psi,\Phi_0)$ and decrease as the inclination angle becomes smaller.
We have higher SNR for the systems with smaller inclination angles, as is shown in Fig.~\ref{snr1}, thus smaller errors for parameters $(M, m_p, a,\Delta \Omega)$.
For parameters $(\iota,d_L,\psi,\Phi_0)$, because of the degeneracy in the small inclination, we get worse parameter estimation in spite of larger SNR.
Comparing the errors with and without higher harmonics, we find that
including higher harmonics can reduce the errors of parameters $(\iota,d_L,\psi,\Phi_0)$, and the improvement induced by higher harmonics depends on the inclination angle.
The smaller the inclination angle is, the bigger improvement we have in parameters $(\iota,d_L,\psi,\Phi_0)$.
Especially, the improvement in parameters $(\iota,d_L,\psi,\Phi_0)$ can reach about five to six orders of magnitude for the small inclination angle around zero.
When the inclination angle $\iota \gtrsim 1.0$, the influence of including higher harmonics on parameters $(\iota,d_L,\psi,\Phi_0)$ can be neglectable.
For the intrinsic parameters $(M, m_p, a)$ and the source localization $\Delta \Omega$, including higher harmonics nearly do not influence the errors of parameters whatever the inclination angle is. 
The reason is that the source localization $\Delta \Omega$ for long-inspiral waveform is mainly dependent on the Doppler effect caused by the motion of the LISA constellation from Eq.~\eqref{doppler}, which is independent of inclination \cite{Zhang:2020drf, Zhang:2020hyx}. 
For the intrinsic parameters $(M, m_p, a)$, these intrinsic parameters are mainly determined by the GW phase as well as the SNR of the signal, and degeneracy caused by the inclination angle has no influence on errors of the intrinsic parameters.
The slight improvement mainly comes from the increasing SNR by including higher harmonics as seen in Fig.~\ref{snr1}.

\begin{figure*}[htp]
  \centering
  \includegraphics[width=0.9\textwidth]{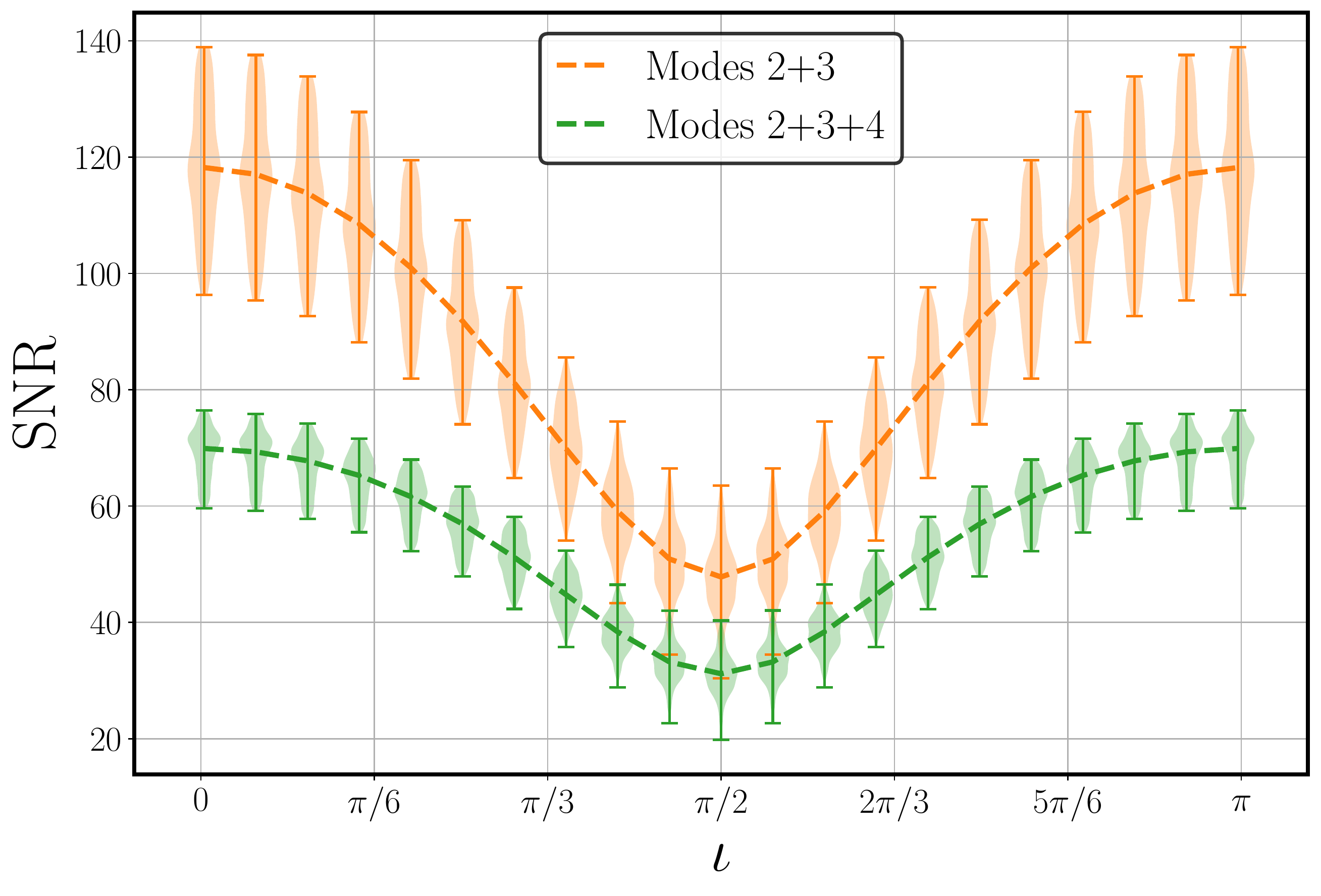}
  \caption{The SNR as functions of inclination angle.
      The orange dashed line represents the medians of SNR of the samples in which the GW waveform includes higher harmonics modes $(l,m)=(2,2),(2,1),(3,3)$, while the green dash line represents the GWs including higher harmonics modes $(l,m)=(2,2),(2,1),(3,3),(3,2),(4,4)$.
  }
  \label{snr2}
\end{figure*}

\begin{figure*}[htp]
  \centering
  \includegraphics[width=0.8\textwidth]{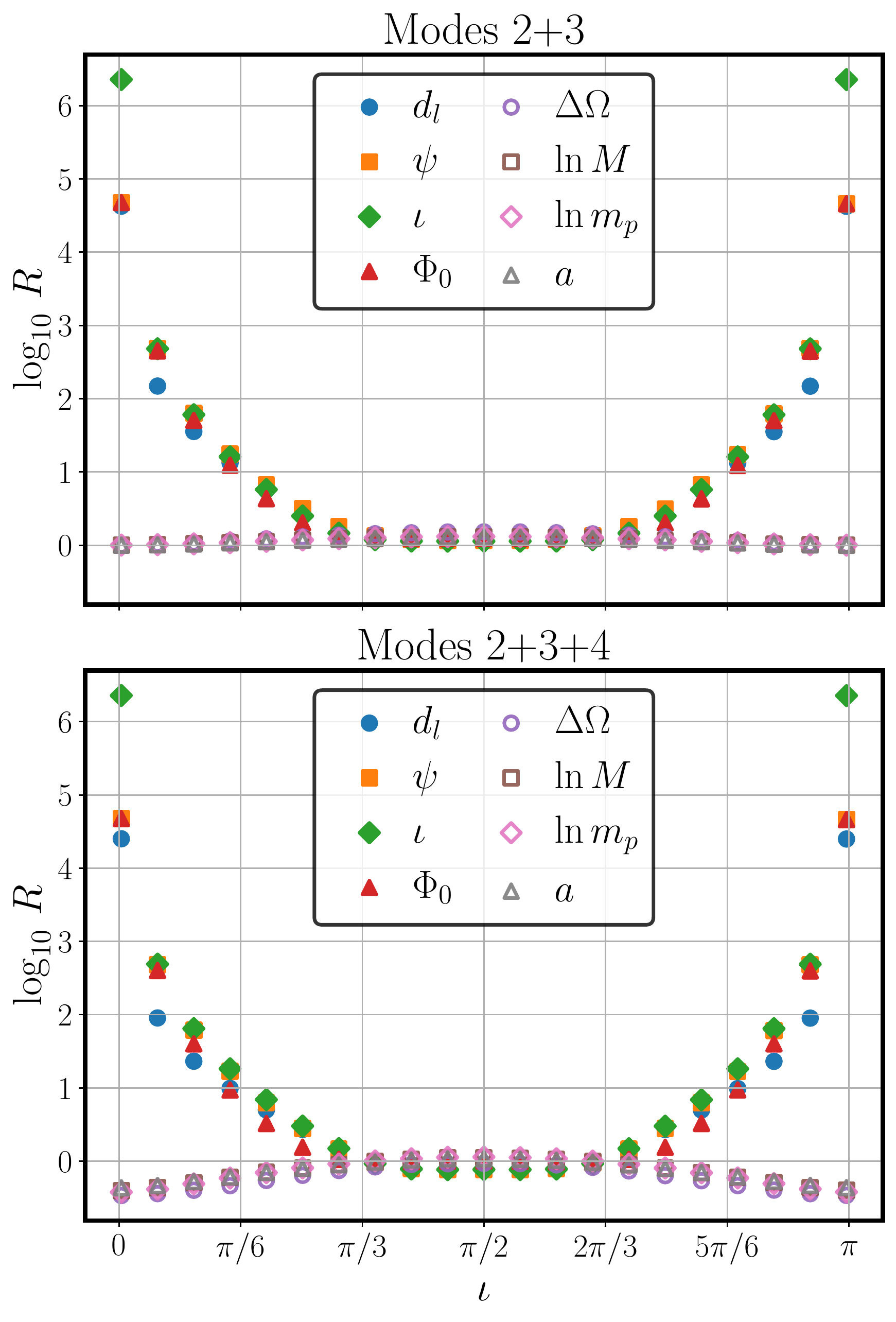}
  \caption{
 The ratios $R$ of improvement induced by including higher harmonic signals as functions of inclination angle.
  The top figure represents the GW waveform including higher harmonics modes $(l,m)=(2,2),(2,1),(3,3)$ and the bottom figure represents the GW waveform including higher harmonics modes $((2,2),(2,1),(3,3),(3,2),(4,4))$.
  }
  \label{FIMresult2}
\end{figure*}

For each inclination angle, we define the ratio $R$ to show the improvement induced by including higher harmonic signals in that orientation,
\begin{equation}
R=\frac{\sigma^{(2,2)}_{\xi_i}}{\sigma^{(2,2)+(2,1)+(3,3)+\cdot\cdot\cdot}_{\xi_i}},
\end{equation}
where $\sigma^{(2,2)}_{\xi_i}$ represents the parameter $\xi_i$ error calculated by using the GW waveform mode $(2,2)$ without higher harmonics, while $\sigma^{(2,2)+(2,1)+(3,3)+\cdot\cdot\cdot}_{\xi_i}$ denotes the parameter $\xi_i$ error estimated including higher harmonics and the number of higher harmonics depends on the situation.
If $R>1$, there is an improvement in the relevant parameter.
A larger $R$ indicates a tighter constraint, and hence a larger improvement.
We also analyze the effect of the number of higher-order modes on parameter estimation.
Figures \ref{snr2} and \ref{FIMresult2} respectively give the medians of SNR and the parameter estimation errors improvement for each inclination value when we use GWs including higher harmonics modes $(l,m)=(2,2),(2,1),(3,3)$, as well as the GWs including higher harmonics modes $(l,m)=(2,2),(2,1),(3,3), (3,2),(4,4)$.
We can see that including more number of higher harmonics can not significantly improve the errors of parameters $(\iota,d_L,\psi,\Phi_0)$ compared with the results from including higher harmonics modes $(l,m)=(2,2),(2,1),(3,3)$.
In contrast, more higher harmonics can even worsen the measurement of source parameters.
The reason is that once the degeneracy is broken, more number of higher harmonics can not break the degeneracy again, and the influence of more higher harmonics on the estimation of the parameters will not be significant but only depends on the SNR as seen in Fig.~\ref{snr2}.
So adding just a few higher harmonics is enough to break the degeneracy and reduce the parameter errors for $(\iota,d_L,\psi,\Phi_0)$ in small inclination for EMRIs.
It is also advantageous for us because adding a large number of higher harmonics can greatly increase our computational cost.

\begin{figure*}[htp]
  \centering
\includegraphics[width=0.93\textwidth]{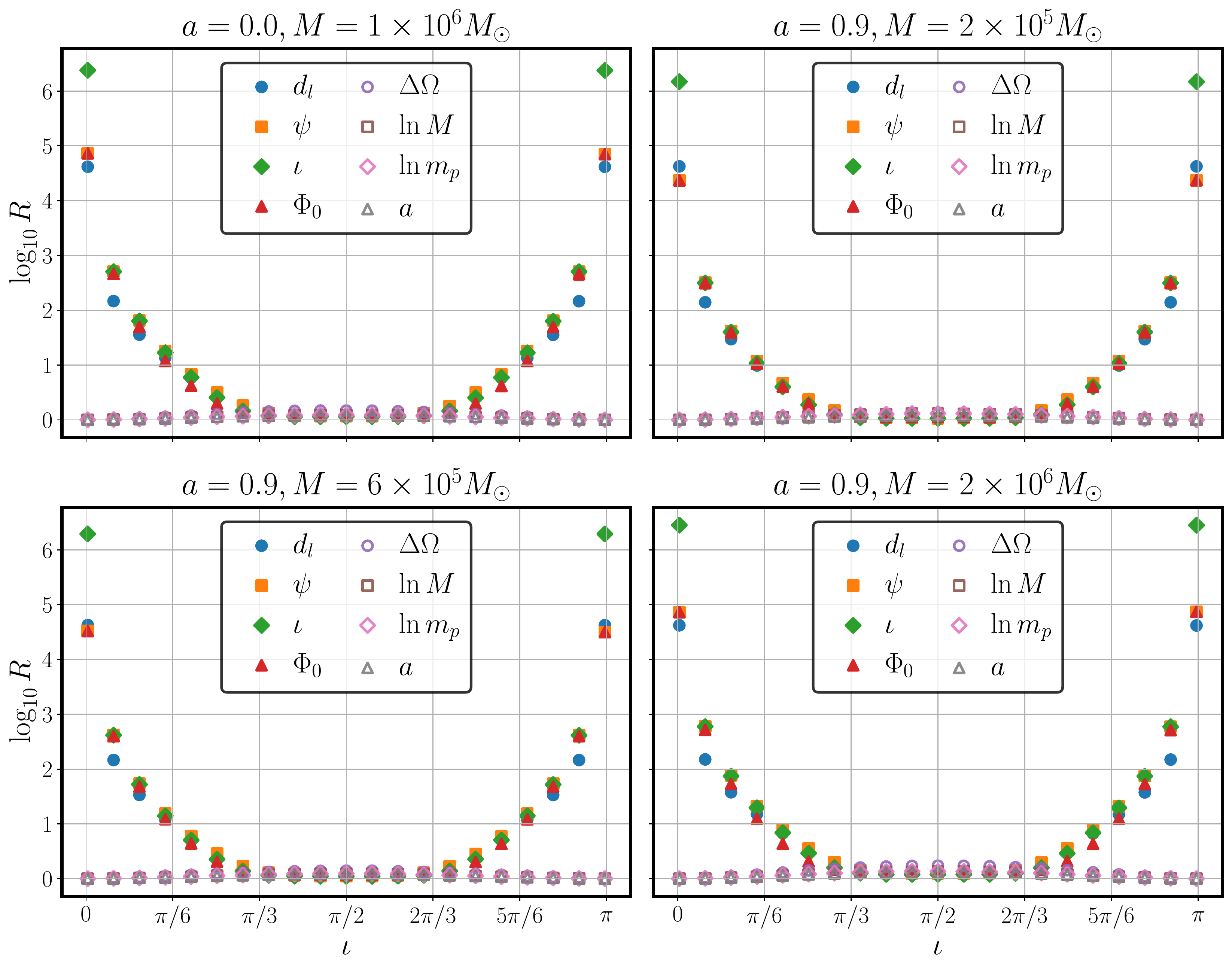}
  \caption{
  The ratios $R$ of improvement induced by including higher harmonic signals $(l,m)=(2,2),(2,1),(3,3)$  as functions of inclination angle.
 The top left figure represents the EMRI system I with $a=0$, $M=10^6~M_{\odot}$, $m_p=10~M_{\odot}$, the top right figure represents the EMRI system II with $a=0.9$, $M=2\times10^5~M_{\odot}$, $m_p=10~M_{\odot}$, the bottom left figure represents the EMRI system III with $a=0.9$, $M=6\times10^5~M_{\odot}$, $m_p=10~M_{\odot}$, and the bottom right figure represents the EMRI system IV with $a=0.9$, $M=2\times10^6~M_{\odot}$, $m_p=10~M_{\odot}$.
  }
  \label{FIMresult3}
\end{figure*}

Furthermore, we give the parameter errors for other kinds of EMRI configurations. 
In Fig.~\ref{FIMresult3}, we give the influence of higher harmonics for the system with different spin $a=0$ (EMRI I), different mass $M=2\times 10^5~M_{\odot}$ (EMRI II), different mass $M=6\times 10^5~M_{\odot}$ (EMRI III) and different mass $M=2\times 10^6~M_{\odot}$ (EMRI IV).
The higher harmonics can greatly improve the errors in parameters $(\iota,d_L,\psi,\Phi_0)$ for small inclination and nearly do not influence the parameters $(M, m_p, a,\Delta \Omega)$ whatever the inclination angle is.
The results are consistent with our previous analysis and independent of the mass or spin of the EMRI system.

\section{Conclusion}
\label{conclusion}
We analyze the influence of higher harmonics on the errors of source parameters in EMRIs.
For the face-on EMRIs, i.e. the inclination angle is small, there are the distance-inclination degeneracy and the initial phase-polarization angle degeneracy and the contribution from the higher order modes can help break some of the degeneracies, thus can significantly reduce the error of measuring the exterior parameters such as inclination angle $\iota$, the luminosity distance $d_L$, the polarization angle $\psi$, and the initial phase $\Phi_0$, except for source localization $\Delta \Omega$.
As the inclination angle increases above $1.0$, the distance-inclination degeneracy and the initial phase-polarization angle degeneracy can be broken naturally even without higher-order modes, so the contribution from higher harmonics becomes insignificant and nearly does not influence the errors of parameters.
For the intrinsic parameters $(M, m_p, a)$ and the source localization $\Delta \Omega$, 
including higher harmonics nearly do not influence the errors of parameters whatever the inclination angle is. 
The errors of source localization $\Delta \Omega$ determined by the Doppler effect, the intrinsic parameters $(M, m_p, a)$ determined by the GW phase as well as the SNR of the signal, are independent of the inclination angle and there are no degeneracies among these parameters even in the small inclination.
We also analyze the effect of the number of higher-order modes on parameter estimation.
Including more higher-order modes can not significantly reduce the errors of parameters $(\iota,d_L,\psi,\Phi_0)$ compared with the results from including higher harmonics modes $(l,m)=(2,2),(2,1),(3,3)$.
Once the degeneracy is broken, more number of higher harmonics can not break the degeneracy again and is insignificant in the improvement of the parameter estimation.
Furthermore, we give the influence of higher harmonics on parameter errors for the system with a different spin and different masses.
The higher harmonics can greatly reduce the errors in parameters $(\iota,d_L,\psi,\Phi_0)$ for small inclination and nearly do not influence the parameters $(M, m_p, a,\Delta \Omega)$ whatever the inclination angle is.
This conclusion is independent of the mass or spin of the EMRI system.

\appendix

\begin{acknowledgments}
We thank the referee for the helpful comments. This work was supported by the China Postdoctoral Science Foundation (2021TQ0018) and the National Natural Science Foundation of China (11975027, 11991053).
\end{acknowledgments}

% \bibliographystyle{apsrev4-2f}
% \bibliography{HigerModes}


\end{document}